\documentclass[]{interact}

\usepackage[T1]{fontenc}
\usepackage{lmodern}
\usepackage{microtype}

\usepackage{epstopdf}
\usepackage[caption=false]{subfig}
\usepackage[numbers,sort&compress]{natbib}
\bibpunct[, ]{[}{]}{,}{n}{,}{,}
\renewcommand\bibfont{\fontsize{10}{12}\selectfont}
\makeatletter
\def\NAT@def@citea{\def\@citea{\NAT@separator}}
\makeatother

\usepackage{amsmath, amssymb, amsthm}
\usepackage{bm}
\usepackage{booktabs}
\usepackage{graphicx}
\usepackage{float}
\usepackage{xcolor}
\usepackage{hyperref}
\usepackage{algorithm}
\usepackage{algorithmic}

\theoremstyle{plain}

\theoremstyle{definition}

\newcommand{\museg}{\mu^{(\text{seg})}}
\newcommand{\muint}{\mu^{(\text{int})}}
\newcommand{\betaseg}{\bm{\beta}^{(\text{seg})}}
\newcommand{\betaint}{\bm{\beta}^{(\text{int})}}
\newcommand{\Xseg}{\bm{X}^{(\text{seg})}}
\newcommand{\Xint}{\bm{X}^{(\text{int})}}
\newcommand{\Normal}{\mathcal{N}}
\newcommand{\Poisson}{\text{Poisson}}
\newcommand{\NB}{\text{NB}}
\newcommand{\NSeg}{8169}
\newcommand{\NInt}{8398}

\makeatletter
\newcommand{\TableCaptionBottom}[2]{%
  \refstepcounter{table}%
  \label{#1}%
  \par\vspace{4pt}%
  {\small\noindent\textbf{Table~\thetable.} #2\par}%
}
\makeatother

\newcommand{\SafeIncludeGraphics}[2][]{%
  \IfFileExists{#2}{\includegraphics[#1]{#2}}{%
    \fbox{\begin{minipage}{0.86\textwidth}\centering
    Missing figure file: \texttt{\detokenize{#2}}
    \end{minipage}}%
  }%
}

\begin{document}

\title{Bayesian Node--Edge Modeling of Road Crashes in Central Bogotá\\
\large Modelación bayesiana nodo--arista de siniestros viales en el centro de Bogotá}

\author{
  \name{Danna Lesley Cruz Reyes\textsuperscript{a}
        and Cristian Harvey Ardila Bolívar\textsuperscript{a}\thanks{
          Corresponding author: Danna Lesley Cruz Reyes.
          Email: dlcruzr@unal.edu.co.
          ORCID: \url{https://orcid.org/0000-0002-5977-8162}.}}
  \affil{\textsuperscript{a}Departamento de Estadística, Facultad de Ciencias,
         Universidad Nacional de Colombia -- Sede Bogotá,
         Bogotá D.C., Colombia.}
}

\maketitle

\begin{abstract}
\textbf{Introduction:} Crash counts on road segments and intersections exhibit
different exposure and connectivity patterns that conventional analyses may obscure.
\textbf{Methodology:} A Bayesian negative-binomial node--edge model was fitted to
8,169 road segments and 8,398 intersections in six central districts of Bogotá. Separate predictors represented
road hierarchy, pavement, speed, signalization, intersection configuration, and land-use
treatment. Model performance was examined through predictive summaries and spatial
diagnostics, while computational details are reported in the appendix.
\textbf{Results:} Intersections with at least four incident segments and higher maximum
incident speeds had higher expected crash counts. Land-use treatment and signalized-access
intensity also showed posterior associations. Road hierarchy and signalization were the
clearest segment-level factors; however, this component had weak raw-scale predictive
performance and numerical uncertainty for some pavement categories.
\textbf{Conclusion:} Treating intersections and segments as distinct network elements
provides an interpretable baseline for urban crash analysis, but the segment results
require cautious interpretation and motivate spatially structured extensions.
\end{abstract}


\section*{Resumen}
\textbf{Introducción:} Los conteos de accidentes en segmentos e intersecciones presentan
patrones distintos de exposición y conectividad que los análisis convencionales pueden
ocultar. \textbf{Metodología:} Se ajustó un modelo bayesiano binomial negativo nodo--arista
a 8.169 segmentos y 8.398 intersecciones de seis localidades centrales de Bogotá. Los predictores separados representaron
jerarquía vial, superficie, velocidad, semaforización, configuración de las intersecciones
y tratamiento de uso del suelo. El desempeño se evaluó mediante resúmenes predictivos y
diagnósticos espaciales, mientras los detalles computacionales se presentan en el anexo.
\textbf{Resultados:} Las intersecciones con al menos cuatro segmentos
incidentes y mayores velocidades máximas presentaron conteos esperados más altos. El
tratamiento del suelo y la intensidad de accesos semaforizados también mostraron
asociaciones posteriores. En los segmentos, la jerarquía vial y la semaforización fueron
los factores más claros, aunque el desempeño predictivo en escala original fue débil.
\textbf{Conclusión:} La distinción entre nodos y aristas ofrece una base interpretable,
pero los resultados de segmentos requieren cautela y motivan extensiones espaciales.

\noindent\textbf{Palabras clave:} inferencia bayesiana; seguridad vial; accidentes de tránsito;
redes urbanas; regresión binomial negativa; análisis espacial.

\section{Introduction}
\label{sec:intro}

Road crashes are not distributed uniformly across a city. Some occur along high-capacity corridors, while others concentrate at intersections where several traffic movements meet. Treating all locations as equivalent can hide these differences. A network representation is useful in this setting because it distinguishes road segments from intersection nodes and retains the connections between them \cite{ZengHuang2014}.

Crash frequency also poses a statistical difficulty. Counts are nonnegative integers and usually vary more than a Poisson model allows. In addition, exposure is not defined in the same way for every network element: a road segment has a measurable length, whereas an intersection is a point at which the number and characteristics of incident roads are more informative. These differences motivate separate predictors for segments and intersections within a common count-data framework.

This study examines road-crash patterns in six central districts of Bogotá using a Bayesian negative-binomial model. The segment component includes road hierarchy, pavement, lane count, speed limit, length, and signalization. The intersection component considers node degree, speed limits on incident roads, land-use treatment, and the number of signalized approaches. The analysis is intended as an interpretable baseline rather than a complete spatial prediction model.

The objective is to determine which road and intersection characteristics are associated with the observed outcomes and to assess whether the node--edge distinction is useful in practice. The article first describes the study network and its spatial pattern, then presents the model and results, and finally discusses their use and limitations for road-safety assessment.

\section{Theoretical Framework}
\label{sec:models}

\subsection{Notation}
\label{subsec:notation}

Let $G=(V,E)$ denote an urban road network, where $E=\{1,\dots,n^{(\text{seg})}\}$ is the set of road segments and $V=\{1,\dots,n^{(\text{int})}\}$ is the set of intersections. For each segment $i\in E$, let $Y_i^{(\text{seg})}$ be the crash count assigned to that segment, and for each intersection $j\in V$, let $Y_j^{(\text{int})}$ be the crash count assigned to that node. Segment and intersection predictors are built from separate design matrices,
\[
\Xseg \in \mathbb{R}^{n^{(\text{seg})} \times p_{\text{seg}}},
\qquad
\Xint \in \mathbb{R}^{n^{(\text{int})} \times p_{\text{int}}},
\]
where $p_{\text{seg}}$ and $p_{\text{int}}$ denote the number of columns in the segment and intersection design matrices, respectively, that is, the number of regression terms used in each linear predictor. The corresponding regression vectors are $\betaseg$ and $\betaint$. Segment offsets are defined by segment length, while intersection offsets are fixed at zero.

\subsection{From the Poisson motivation to the negative binomial layer}
\label{subsec:poissonnb}

A natural starting point for count data is the Poisson model,
\[
Y_i \mid \lambda_i \sim \Poisson(\lambda_i),
\qquad
\log \lambda_i = \log(E_i) + x_i^\top\beta.
\]
Under this specification,
\[
\mathbb{E}(Y_i \mid x_i)=\lambda_i,
\qquad
\mathrm{Var}(Y_i \mid x_i)=\lambda_i,
\]
so the conditional mean and variance are equal. For traffic crashes, this restriction is often too strong, since variability usually exceeds what a Poisson law can accommodate.

A standard way to relax that restriction is to introduce multiplicative unobserved heterogeneity,
\[
Y_i \mid \phi_i \sim \Poisson(\mu_i e^{\phi_i}),
\qquad
\mu_i = \exp\!\big(\log(E_i)+x_i^\top\beta\big),
\]
and assume that $e^{\phi_i} \sim \Gamma(a,a)$, where the Gamma distribution is parameterized by shape and rate. Marginalizing over this heterogeneity yields a negative binomial law,
\[
Y_i \mid \beta,\kappa \sim \NB(\mu_i,\kappa),
\qquad \text{with } \kappa=\frac{1}{a},
\]
where $\NB(\mu_i,\kappa)$ denotes the mean--overdispersion parameterization of the negative binomial distribution. Under this parameterization,
\[
\mathbb{E}(Y_i \mid x_i)=\mu_i,
\qquad
\mathrm{Var}(Y_i \mid x_i)=\mu_i + \kappa\mu_i^2.
\]
This theoretical construction justifies the negative binomial layer used throughout the article. In the implemented model, however, the auxiliary Gamma effect is not sampled explicitly. The empirical specification works by applying a marginal negative binomial law, which substantially reduces the use of computational resources.

\subsection{Negative binomial node--edge model}
\label{subsec:modelA}

The implemented model applies a negative binomial specification separately to segments and intersections while preserving the same observational logic in both cases. For segments,
\[
Y_i^{(\text{seg})} \mid \betaseg,\kappa^{(\text{seg})} \sim \NB(\museg_i,\kappa^{(\text{seg})}),
\qquad
\log \museg_i = \log E_i + (\Xseg_i)^\top\betaseg,
\]
and for intersections,
\[
Y_j^{(\text{int})} \mid \betaint,\kappa^{(\text{int})} \sim \NB(\muint_j,\kappa^{(\text{int})}),
\qquad
\log \muint_j = \log E_j + (\Xint_j)^\top\betaint.
\]
Because intersections are treated as nodes rather than linear entities, their exposure is fixed at one, so that $E_j=1$ and the offset is zero. The model therefore captures observed covariates, exposure, and overdispersion, while maintaining a clear distinction between the two types of entities in the road graph.

The prior specification and full computational details are provided in Appendix~\ref{app:technical_details} so that the main text can focus on the road-safety application and interpretation.

\section{Data}
\label{sec:data}

The analysis uses georeferenced crash records from the public road-crash layer of Bogotá's Integrated Mobility Information System (SIMUR), administered by the Bogotá District Mobility Secretariat \cite{SIMURAccidentalidad}, together with an OpenStreetMap representation of the road network. The study area comprises six central districts: Chapinero, Barrios Unidos, Los Mártires, Teusaquillo, La Candelaria, and Santa Fe. The final georeferenced representation integrates graph topology, segment and node geometries, crash outcomes, and the covariates required by the node--edge model. The spatial information is transformed to EPSG:3116 (MAGNA-SIRGAS / Colombia Bogotá zone) to support distance and network operations.

The analysis therefore works with two types of units. The first is the road segment, corresponding to an edge of the road graph. The second is the intersection, corresponding to a node of the graph. The final working dataset contains \NSeg{} segments and \NInt{} intersections.

In the available preprocessing workflow, each georeferenced crash point is assigned to
its nearest OpenStreetMap segment and segment counts are stored as $\log(1+Y)$. The
segment response used in the negative-binomial model is recovered with the inverse
transformation $\exp(Y)-1$. The available intersection response was not obtained through
an independent point-to-intersection assignment. Instead, it was constructed by summing
the stored $\log(1+Y)$ values of the edges incident to each node and rounding the result
to a nonnegative integer before model fitting. This construction is reported explicitly
because it affects the interpretation of the intersection component and should be
replaced by direct point-based intersection counts in a future data-processing revision.

Crash counts are represented by two response vectors,
\[
Y_i^{(\text{seg})} \in \{0,1,2,\dots\},
\qquad i=1,\dots,n^{(\text{seg})},
\]
for segments, and
\[
Y_j^{(\text{int})} \in \{0,1,2,\dots\},
\qquad j=1,\dots,n^{(\text{int})},
\]
for intersections. For segments, exposure is defined by physical length,
\[
E_i = \text{length}_i,
\qquad
\log E_i = \log(\text{length}_i),
\]
whereas for intersections exposure is fixed at one,
\[
E_j = 1,
\qquad
\log E_j = 0.
\]
This distinction reflects the difference between linear entities with measurable extent and intersection nodes.

\begin{table}[htbp]
\centering
\begin{tabular}{lrrrrrr}
\toprule
Entity & $n$ & Mean & SD & Min & Median & Max \\
\midrule
Road segments & 8169 & 11.726 & 27.550 & 0 & 3 & 815 \\
Intersections & 8398 & 2.874 & 3.150 & 0 & 2 & 23 \\
\bottomrule
\end{tabular}
\TableCaptionBottom{tab:descriptiva}{Descriptive statistics of crash counts by entity type.}
\end{table}

\subsection{Covariates}
\label{subsec:covariates}

Segment covariates summarize the geometric, operational, and control-related characteristics of the roadway. In the final implementation, the segment predictor is built from the standardized lane count \texttt{carriles\_sc}, the road-type factor \texttt{tipo\_via}, the pavement-surface factor \texttt{superficie}, the standardized maximum-speed variable \texttt{vel\_max\_sc}, and the binary signalization variable \texttt{via\_semaforizada\_bin}. Lane counts are recovered from the available lane variables, road class is collapsed into a reduced hierarchy, pavement surface is grouped into a smaller set of functional categories, maximum speed is extracted numerically from speed-limit fields, and signalized segments are coded as a binary indicator.

Missing values are handled through variable-specific rules as follows. Lane counts are imputed using medians within road-type groups, with a global fallback. Surface is imputed using the modal category within road-type groups, again with a global fallback. Maximum speed is imputed using the validated hierarchy adopted in the final implementation: arterial and secondary/collector roads receive a default of 50, local and service roads receive a default of 30, and the global modal speed is used when road-type information is missing or uninformative. This rule is substantively preferable to a generic fill-in because it preserves the basic road hierarchy built into the data preparation process.

Intersection covariates are derived from the node--edge structure of the network and from georeferenced urban covariates. The final predictor uses the indicator \texttt{no\_leg4} for intersections with degree at least four,
\[
\texttt{no\_leg4}_j = \mathbf{1}\{\deg(j)\geq 4\},
\]
together with the standardized variables \texttt{major\_sl\_sc} and \texttt{minor\_sl\_sc}, constructed from the highest and second-highest speed limits among the segments incident to each node. The intersection design also includes the land-use treatment factor \texttt{trat\_nombre\_tratamiento} and the standardized signalized-access intensity \texttt{sem\_accesos\_sc}. These variables summarize the operational configuration of each intersection from the perspective of the surrounding corridor segments and the urban conditions associated with the node.

\begin{table}[htbp]
\centering
\small
\begin{tabular}{llrr}
\toprule
Entity & Variable & Before (\%) & After (\%) \\
\midrule
Road segments & Lanes & 81.6 & 100.0 \\
Road segments & Surface & 92.8 & 100.0 \\
Road segments & Maximum speed & 28.7 & 100.0 \\
Road segments & Signalized segment & 100.0 & 100.0 \\
Intersections & Four or more incident road segments & 100.0 & 100.0 \\
Intersections & Highest incident speed & 100.0 & 100.0 \\
Intersections & Second-highest incident speed & 100.0 & 100.0 \\
Intersections & Land-use treatment & 89.4 & 100.0 \\
Intersections & Signalized-access intensity & 100.0 & 100.0 \\
\bottomrule
\end{tabular}
\TableCaptionBottom{tab:cobertura}{Variable coverage before and after imputation.}
\end{table}

\subsection{Descriptive spatial context}
\label{subsec:descriptive_spatial}

\begin{figure}
\centering
\SafeIncludeGraphics[width=0.92\textwidth]{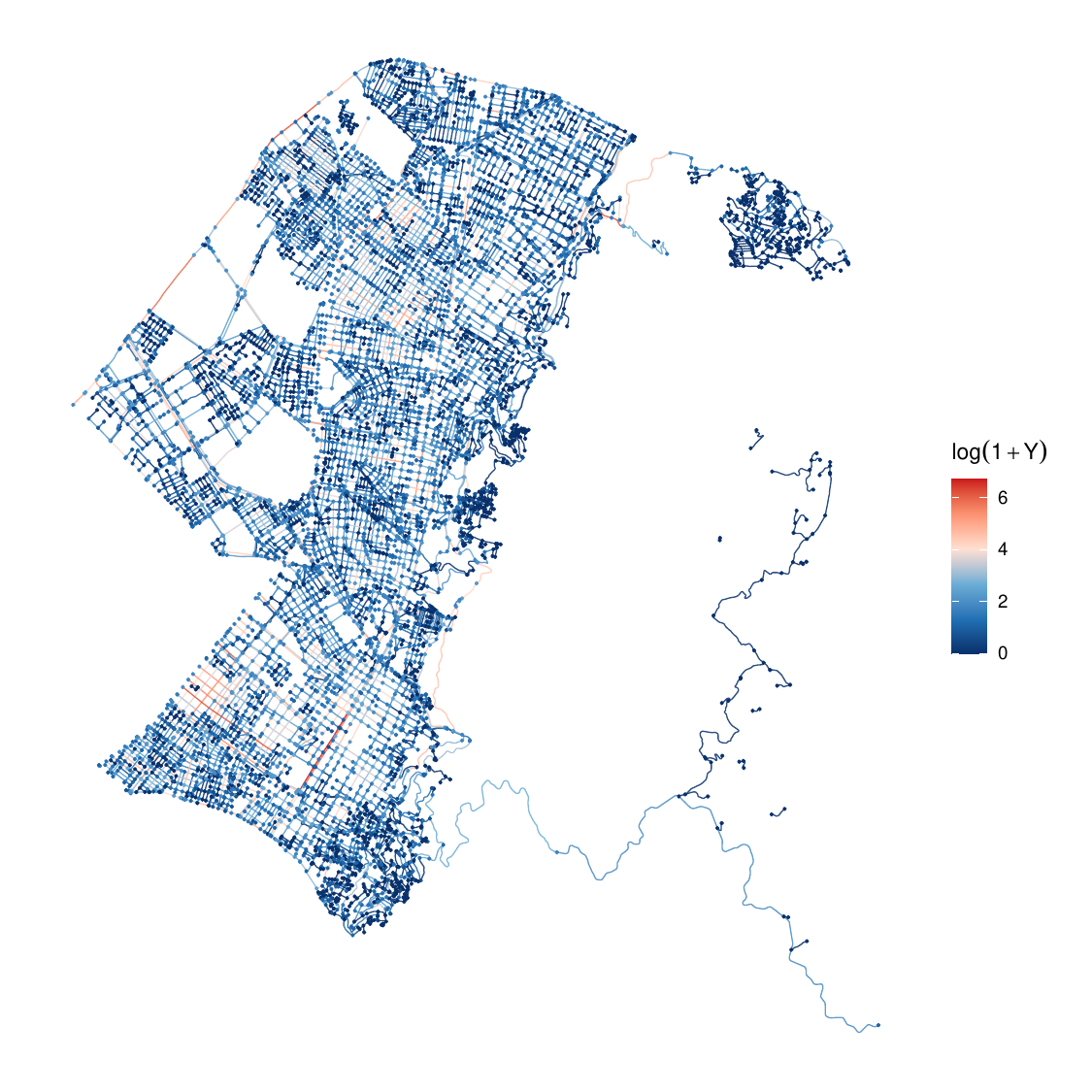}
\caption{Observed crash counts on the Bogotá road network. Colors represent $\log(1+Y)$ on a common scale for segments and intersections.}
\label{fig:mapa_obs_descriptivo}
\end{figure}

The descriptive spatial analysis is used to examine the empirical distribution of crashes before presenting the inferential results of the Bayesian model. Figure~\ref{fig:mapa_obs_descriptivo} shows the observed crash counts on the road network. The display uses the transformed scale $\log(1+Y)$ for both segments and intersections, which makes the spatial concentration of crashes visible while limiting the dominance of extreme counts.

The descriptive spatial association was summarized through a global Moran index computed separately on the two supports of the network \cite{BivandWong2018}. For road segments, adjacency is defined by shared endpoints between edges. For intersections, adjacency is defined by direct connection through a road segment. A mixed Moran index over points and lines is not reported because segments and intersections are different spatial supports. The expectation and variance reported in Table~\ref{tab:moran_descriptivo} correspond to the reference distribution of Moran's I under spatial randomness.

\begin{table}[htbp]
\centering
\small
\begin{tabular}{lrrrrr}
\toprule
Support & Moran I & Expected $I$ & Var($I$) & Z & $p$-value \\
\midrule
Road segments / edge adjacency & 0.2027 & -0.0001 & 0.000100 & 20.235 & $<0.001$ \\
Intersections / node adjacency & 0.6887 & -0.0001 & 0.000147 & 56.719 & $<0.001$ \\
\bottomrule
\end{tabular}
\TableCaptionBottom{tab:moran_descriptivo}{Global Moran index for observed crash counts on the two network supports.}
\end{table}

\section{Methodology}
\label{sec:inference}

\subsection{Implementation of the theoretical framework into the data}
\label{subsec:mapping}

To implement the theoretical model in our data, the following workflow was used. Let
\[
y^{(\text{seg})}=(Y_1^{(\text{seg})},\dots,Y_{n^{(\text{seg})}}^{(\text{seg})})^\top,
\qquad
y^{(\text{int})}=(Y_1^{(\text{int})},\dots,Y_{n^{(\text{int})}}^{(\text{int})})^\top
\]
denote the observed crash-count vectors for segments and intersections. Let
\[
\Xseg \in \mathbb{R}^{n^{(\text{seg})}\times p_{\text{seg}}},
\qquad
\Xint \in \mathbb{R}^{n^{(\text{int})}\times p_{\text{int}}}
\]
be the corresponding design matrices, and let
\[
o^{(\text{seg})}=\log E^{(\text{seg})},
\qquad
o^{(\text{int})}=\log E^{(\text{int})}
\]
be the offset vectors. In the present application, $o^{(\text{seg})}$ is the log-length vector for segments and $o^{(\text{int})}$ is the zero vector for intersections.

The design matrices are built separately for the two types of entities. Segment candidate terms are \texttt{carriles\_sc}, \texttt{tipo\_via}, \texttt{superficie}, \texttt{vel\_max\_sc}, and \texttt{via\_semaforizada\_bin}; intersection candidate terms are \texttt{no\_leg4}, \texttt{major\_sl\_sc}, \texttt{minor\_sl\_sc}, \texttt{trat\_nombre\_tratamiento}, and \texttt{sem\_accesos\_sc}. This separate construction is methodologically important because segments and intersections are treated as different mathematical elements in the road network.

The empirical implementation therefore uses the following objects: response vectors, offsets, design matrices, and georeferenced network layers for visualization of observed counts, fitted values, and model residuals.

\subsection{Bayesian estimation and model assessment}
\label{subsec:estimation}

The two negative-binomial components were estimated in \textsf{R} using a custom Bayesian algorithm. Posterior means and 95\% credible intervals were obtained for the regression coefficients and converted to relative rates through exponentiation. Model behavior was assessed separately for segments and intersections using mean absolute deviation, mean squared prediction error, a raw-scale predictive $R^2$, and spatial maps of fitted values and absolute errors. Complete prior distributions, simulation settings, acceptance rates, effective sample sizes, autocorrelation summaries, and deviance-based diagnostics are reported in Appendix~\ref{app:technical_details}.

\subsection{Posterior summaries and model evaluation}
\label{subsec:evaluation}

Posterior means are used to construct fitted values for segments and intersections,
\[
\widehat{\museg}_i = \mathbb{E}(\museg_i \mid \text{data}),
\qquad
\widehat{\muint}_j = \mathbb{E}(\muint_j \mid \text{data}),
\]
together with posterior summaries for the regression coefficients and overdispersion parameters.

On each support, predictive accuracy is evaluated on the original count scale using
\[
R^2 = 1-\frac{\sum_i (Y_i-\widehat{\mu}_i)^2}
{\sum_i (Y_i-\bar Y)^2},
\qquad
\mathrm{MAD}=\frac{1}{n}\sum_i |Y_i-\widehat{\mu}_i|,
\qquad
\mathrm{MSPE}=\frac{1}{n}\sum_i (Y_i-\widehat{\mu}_i)^2.
\]
Thus, the reported $R^2$ is a raw-scale predictive summary and may take negative values when the fitted values perform worse than the sample mean on that scale.

Predictive performance is assessed separately for segments and intersections and then aggregated to the network level through weighted averages of the form
\[
W(M)=\frac{n^{(\text{seg})} M_{\text{seg}} + n^{(\text{int})} M_{\text{int}}}{n^{(\text{seg})}+n^{(\text{int})}},
\]
where $M$ denotes a metric computed separately for the two types of entities. In particular, the implementation reports weighted counterparts of $R^2$, MAD, and MSPE, following the same node--edge aggregation logic. These weighted values should be interpreted as descriptive network-level summaries of support-specific metrics, not as pooled metrics computed from a single combined response vector.

\section{Results}
\label{sec:results}

\subsection{Road and intersection factors associated with crashes}
\label{subsec:parametros}

Tables~\ref{tab:modelA_segments_parameters} and~\ref{tab:modelA_intersections_parameters} report posterior means, 95\% credible intervals, and relative rates for the two parts of the model.

Road hierarchy and signalization provide the clearest segment-level contrasts. Relative to the reference road category, the estimated rate is 2.889 times higher for secondary or collector roads and 3.585 times higher for primary arterial roads. The estimated relative rate for signalized segments is 1.896. In contrast, lane count has almost no estimated association, and the interval for maximum speed includes zero on the coefficient scale. Although some pavement categories show large contrasts, their numerical diagnostics are weak; those coefficients are therefore not treated as firm findings.

\begin{table}[htbp]
\centering
\small
\resizebox{\textwidth}{!}{%
\begin{tabular}{lrrrr}
\toprule
Parameter & Mean & 95\% CrI & Mean RR & 95\% RR CrI \\
\midrule
Intercept & -2.980 & [-3.040, -2.916] & 0.051 & [0.048, 0.054] \\
Lanes (standardized) & -0.000 & [-0.039, 0.039] & 1.000 & [0.962, 1.040] \\
Secondary/collector road & 1.061 & [0.948, 1.172] & 2.889 & [2.581, 3.228] \\
Primary arterial road & 1.277 & [1.106, 1.458] & 3.585 & [3.022, 4.298] \\
Concrete surface & 0.065 & [-0.089, 0.218] & 1.067 & [0.915, 1.244] \\
Paving-stone surface & 0.731 & [0.393, 1.123] & 2.077 & [1.481, 3.074] \\
Unpaved/earth surface & -2.823 & [-3.531, -2.238] & 0.059 & [0.029, 0.107] \\
Other surface & -2.446 & [-2.841, -2.108] & 0.087 & [0.058, 0.122] \\
Maximum speed (standardized) & 0.025 & [-0.027, 0.074] & 1.025 & [0.974, 1.077] \\
Signalized segment & 0.640 & [0.545, 0.732] & 1.896 & [1.725, 2.079] \\
Overdispersion $\kappa$ & 2.769 & [2.669, 2.869] & --- & --- \\
\bottomrule
\end{tabular}%
}
\TableCaptionBottom{tab:modelA_segments_parameters}{Posterior summaries of the negative-binomial model for road segments with the enriched covariate set.}
\end{table}

The intersection results are easier to interpret. Nodes connected to four or more road segments have an estimated rate 2.422 times that of less connected nodes. A one-standard-deviation increase in the highest speed limit among incident roads is associated with a relative rate of 1.443. The second-highest incident speed has a smaller inverse association (0.939). Differences are also observed across land-use treatments, while each standard-deviation increase in signalized-access intensity is associated with a relative rate of 1.105.

\begin{table}[htbp]
\centering
\small
\resizebox{\textwidth}{!}{%
\begin{tabular}{lrrrr}
\toprule
Parameter & Mean & 95\% CrI & Mean RR & 95\% RR CrI \\
\midrule
Intercept & 0.623 & [0.593, 0.653] & 1.864 & [1.810, 1.921] \\
Four or more incident road segments & 0.885 & [0.831, 0.940] & 2.422 & [2.295, 2.560] \\
Highest incident speed (standardized) & 0.366 & [0.327, 0.406] & 1.443 & [1.387, 1.501] \\
Second-highest incident speed (standardized) & -0.063 & [-0.104, -0.022] & 0.939 & [0.901, 0.978] \\
Land-use treatment: Renewal & 0.410 & [0.370, 0.455] & 1.506 & [1.447, 1.576] \\
Land-use treatment: Conservation & 0.228 & [0.172, 0.287] & 1.257 & [1.188, 1.332] \\
Land-use treatment: Development & -0.216 & [-0.368, -0.065] & 0.806 & [0.692, 0.937] \\
Land-use treatment: Integral improvement & -0.434 & [-0.602, -0.301] & 0.648 & [0.548, 0.740] \\
Signalized-access intensity (standardized) & 0.100 & [0.079, 0.119] & 1.105 & [1.082, 1.127] \\
Overdispersion $\kappa$ & 0.451 & [0.420, 0.483] & --- & --- \\
\bottomrule
\end{tabular}%
}
\TableCaptionBottom{tab:modelA_intersections_parameters}{Posterior summaries of the negative-binomial model for intersections with the enriched covariate set.}
\end{table}

Overdispersion is considerably larger for segments ($\kappa=2.769$) than for intersections ($\kappa=0.451$). This difference is consistent with the broader range of segment counts and the greater variation in segment length and road function.

\subsection{Reliability and scope of interpretation}
\label{subsec:mcmc_results}

The main interpretation is restricted to coefficients with clear credible intervals and comparatively better numerical behavior. Several pavement estimates and the less frequent land-use categories do not meet both conditions. The full simulation diagnostics are reported in Appendix~\ref{app:technical_details}.

\subsection{Predictive performance}
\label{subsec:ajuste}

Predictive performance differs sharply between the two supports (Table~\ref{tab:modelA_fit_metrics}). For intersections, $R^2=0.349$, MAD $=1.898$, and MSPE $=6.461$. For segments, $R^2=-0.476$, MAD $=14.419$, and MSPE $=1119.842$. The negative segment $R^2$ means that, on the original count scale, its squared error exceeds that of using the sample mean as a prediction. The weighted network summary is therefore also slightly negative.

\begin{table}[htbp]
\centering
\small
\begin{tabular}{lrrr}
\toprule
Block & $R^2$ & MAD & MSPE \\
\midrule
Segments & -0.476 & 14.419 & 1119.842 \\
Intersections & 0.349 & 1.898 & 6.461 \\
Weighted & -0.058 & 8.072 & 555.457 \\
\bottomrule
\end{tabular}
\TableCaptionBottom{tab:modelA_fit_metrics}{Goodness-of-fit metrics for the negative-binomial model.}
\end{table}

The segment component should thus be read as an exploratory association model, not as an accurate predictor of individual segments. Extreme counts and variation in segment exposure have a strong effect on its raw-scale errors. The intersection component performs better, although its fit remains moderate rather than conclusive.

\textbf{}

\subsection{Fitted values and spatial diagnostics}
\label{subsec:figures_results}

\begin{figure}
\centering
\SafeIncludeGraphics[width=0.92\textwidth]{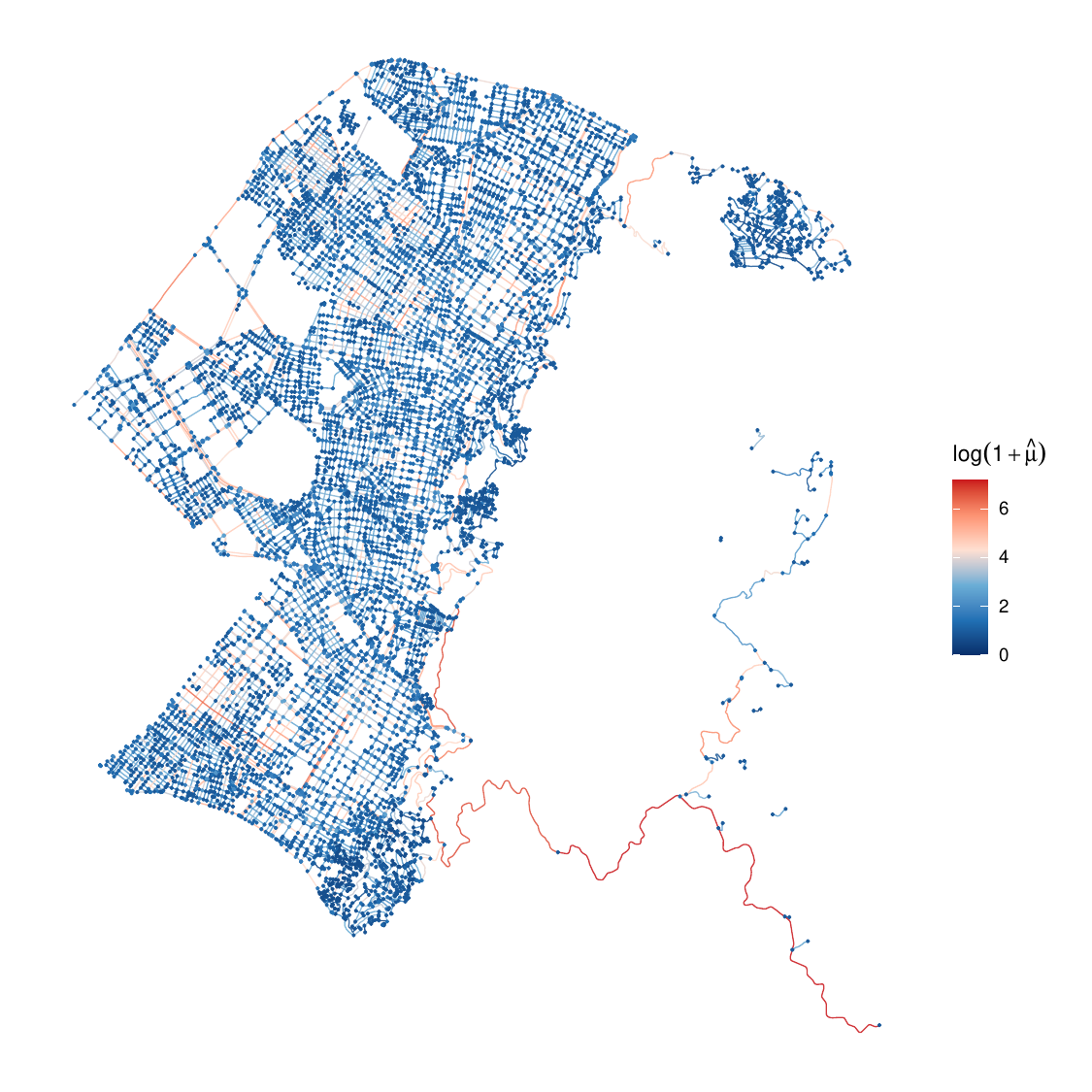}
\caption{Posterior mean crash counts estimated by the negative-binomial model. Colors represent $\log(1+\widehat\mu)$ on a common scale for segments and intersections.}
\label{fig:mapa_est}
\end{figure}

\begin{figure}
\centering
\SafeIncludeGraphics[width=0.92\textwidth]{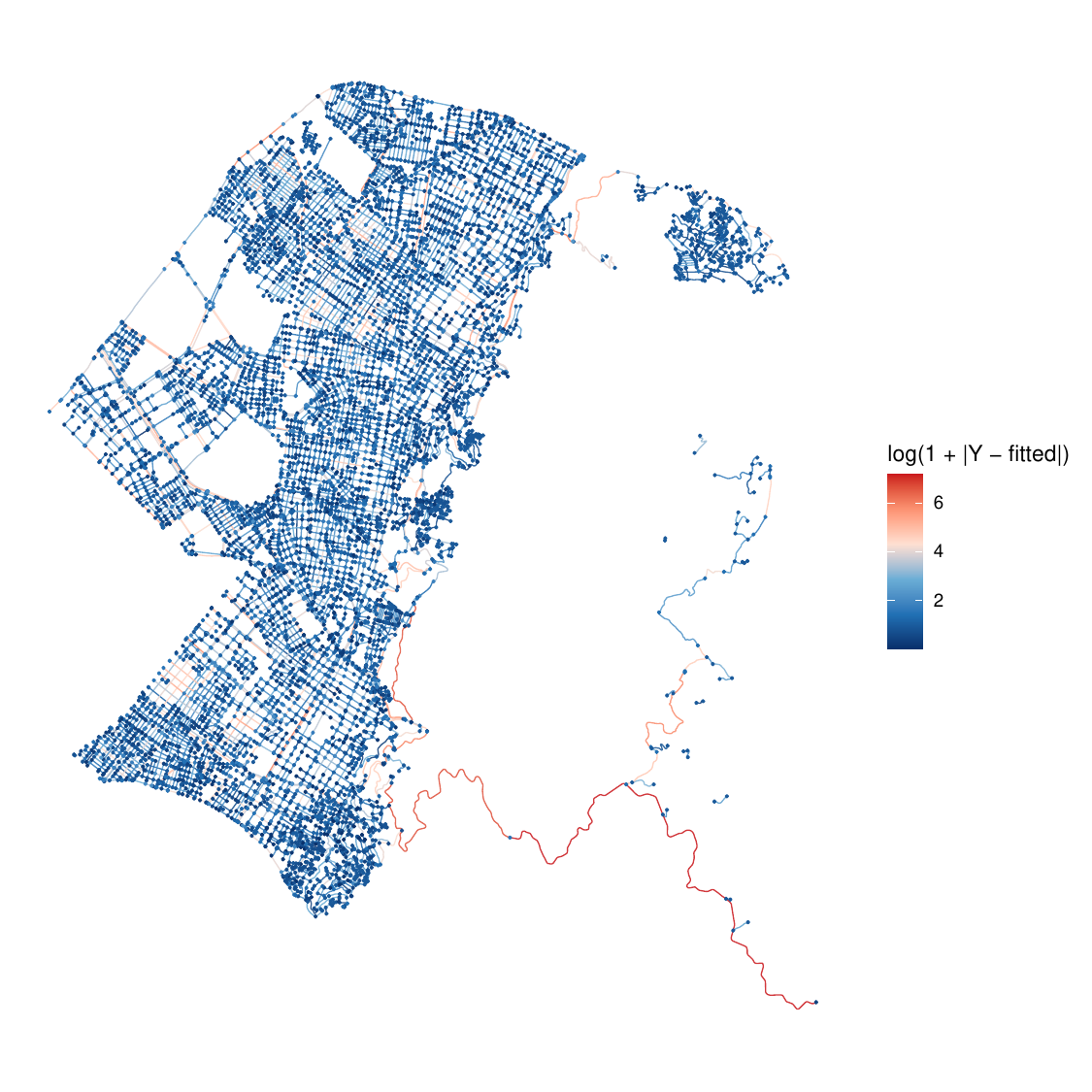}
\caption{Absolute fitting error in log-transformed scale, computed from observed and posterior mean crash counts.}
\label{fig:error_log}
\end{figure}

\begin{figure}
\centering
\SafeIncludeGraphics[width=0.82\textwidth]{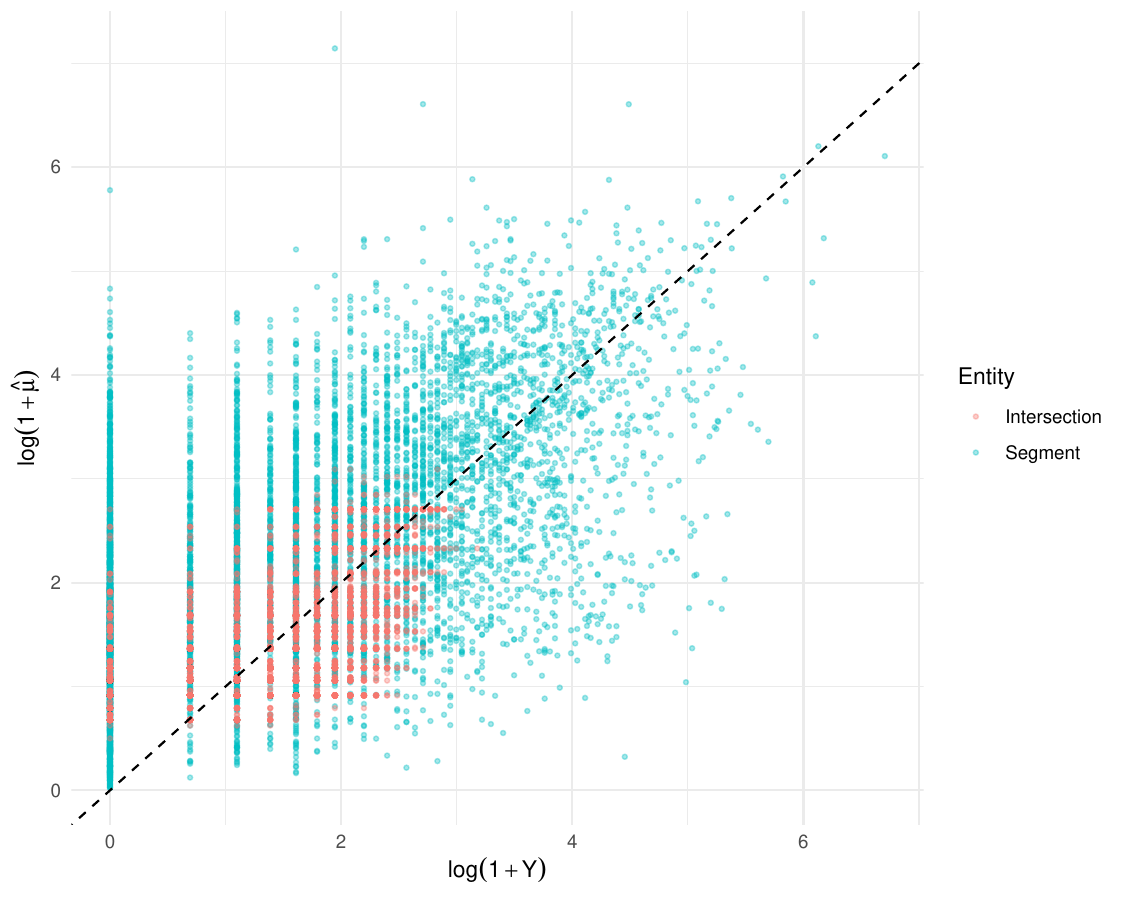}
\caption{Observed versus estimated crash counts in log-transformed scale for segments and intersections.}
\label{fig:scatter_log}
\end{figure}

Figures~\ref{fig:mapa_est}--\ref{fig:scatter_log} show where the model reproduces the broad spatial pattern and where its errors are larger. The maps and scatter plot use a logarithmic transformation so that a small number of extreme segment counts do not dominate the display.

The figures lead to the same conclusion as the numerical metrics: the model separates many low- and high-outcome areas on the transformed scale, but it does not reproduce extreme segment counts well.

\subsection{Road-safety interpretation and practical implications}
\label{subsec:practical_implications}

The intersection results point to a practical screening rule: highly connected nodes, especially those linked to faster roads or several signalized approaches, merit closer examination. This does not by itself identify an unsafe design, but it narrows the set of locations for which field observations, severity records, turning movements, and pedestrian conditions should be reviewed.

For segments, the higher rates estimated for arterial, secondary, collector, and signalized roads probably reflect a combination of road function and unmeasured exposure. A signalized segment should not be interpreted as hazardous merely because it has a signal; signals are commonly installed where traffic and crossing activity are already high. The maps are therefore most useful for identifying corridors that require additional information, rather than for assigning a final risk ranking.

Any intervention decision would need evidence not included here, particularly traffic volume, pedestrian and cyclist flows, crash severity, and a site-level engineering audit. This qualification is especially important for segments because their predictive fit is weak.

\section{Discussion}
\label{sec:discussion}

The variance in both outcomes exceeds what a Poisson specification would accommodate, most notably for road segments. The negative-binomial formulation is therefore reasonable as a first model. Separating nodes and edges is also informative: the two supports have different covariates, exposure definitions, and predictive behavior.

Positive Moran statistics are observed for both supports, with a much stronger value for intersections. Nearby network elements consequently tend to have more similar outcomes than would be expected under spatial randomness. Because the fitted model does not include a spatial random effect, this pattern also signals residual structure that the current covariates may not fully explain.

\subsection{Intersections as conflict locations}

The clearest intersection contrast is node degree: locations with four or more incident segments have substantially higher expected outcomes. The positive association with the highest incident speed is also consistent with a more demanding operating environment. Land-use treatment and signalized-access intensity add contextual information, although the uncommon treatment categories are estimated less precisely. These results describe associations; they do not establish that any one feature causes crashes.

\subsection{Road hierarchy, signalization, and exposure}

Road hierarchy and signalization are the most consistent segment-level results. Larger corridors carry more movement and concentrate crossings and junctions, so their higher estimated rates are plausible. However, traffic volume is absent from the model and segment length is only a partial measure of exposure. The road-class coefficients may therefore capture both roadway characteristics and unobserved demand. The weak predictive fit and unstable pavement estimates reinforce the need for a restrained interpretation.

\subsection{Value of the node--edge perspective}

The main value of the node--edge formulation lies in this contrast. A single distributional family can be used for both supports, but the same predictors and exposure assumptions cannot. Intersections produce a clearer pattern, whereas segments retain considerable unexplained variation. Reporting the components separately makes that difference visible instead of hiding it in a pooled analysis.

\medskip
\noindent\textbf{Limitations.}
The study covers only six central districts, so the results should not be generalized to the entire city. The serialized modeling file does not retain crash dates, and the observation period must be recovered from the source extract before submission. Several OpenStreetMap covariates required deterministic imputation, and traffic volume was unavailable. The intersection response was constructed from incident-edge values rather than from an independent assignment of crash points to nodes, which limits its interpretation as a direct count. Finally, inference is based on one MCMC chain; several coefficients have low effective sample sizes or high autocorrelation. These issues are most consequential for pavement and uncommon land-use categories and should be addressed in a future reconstruction of the data and model.

\section{Conclusion}
\label{sec:conclusion}

The node--edge analysis shows that intersections and road segments should not be treated as interchangeable units. In the six central districts studied, intersections with at least four incident roads and higher incident speeds have larger expected outcomes. Road hierarchy and signalization provide the clearest segment-level contrasts, but the segment model leaves substantial variation unexplained.

The results are best used as an exploratory description of the network. They can help identify intersections and corridors for closer examination, but they are not sufficient for selecting interventions or making causal claims. Direct intersection counts, traffic exposure, crash severity, and spatial residual effects are needed before the model can support stronger site-level decisions.

Despite these limitations, the analysis establishes a reproducible baseline and shows why nodes and edges benefit from separate specifications. Its principal contribution is practical rather than computational: it organizes heterogeneous road elements within one framework while preserving the differences that matter for interpretation.

\section*{Data availability}
The road network data were obtained from OpenStreetMap
(\url{https://www.openstreetmap.org}). Georeferenced crash records were obtained from
the public accidentalidad layer of the Integrated Mobility Information System (SIMUR),
administered by the Bogotá District Mobility Secretariat. The exact observation period
will be reported after verification against the source extract. The information needed to understand the analysis is
reported in this article.

\clearpage
\appendix

\section{Bayesian computation and numerical diagnostics}
\label{app:technical_details}

The regression coefficients were assigned diffuse Gaussian priors,
$\betaseg\sim\Normal(0,100^2I)$ and $\betaint\sim\Normal(0,100^2I)$, while the
overdispersion parameters followed $\Gamma(0.001,0.001)$ distributions. The marginal
negative-binomial likelihood was evaluated directly, without explicitly sampling the
auxiliary Gamma effect.

Posterior simulation used a custom random-walk Metropolis--Hastings algorithm in
\textsf{R}. Regression coefficients were updated by blocks and the overdispersion
parameters on the log scale. The simulation comprised 500,000 iterations, a burn-in
period of 450,000 iterations, and thinning every 10 iterations, producing 5,000 retained
draws. The acceptance rates were 0.196 and 0.273 for the segment regression and
overdispersion blocks, respectively, and 0.026 and 0.459 for the corresponding
intersection blocks.

\begin{table}[H]
\centering
\scriptsize
\resizebox{\textwidth}{!}{%
\begin{tabular}{llrrr}
\toprule
Block & Parameter & ESS & ESS/n & ACF lag 1 \\
\midrule
Segments & Intercept & 611 & 0.122 & 0.671 \\
Segments & Lanes (standardized) & 1931 & 0.386 & 0.291 \\
Segments & Secondary/collector road & 407 & 0.081 & 0.817 \\
Segments & Primary arterial road & 246 & 0.049 & 0.900 \\
Segments & Concrete surface & 351 & 0.070 & 0.869 \\
Segments & Paving-stone surface & 67 & 0.013 & 0.974 \\
Segments & Unpaved/earth surface & 16 & 0.003 & 0.993 \\
Segments & Other surface & 65 & 0.013 & 0.974 \\
Segments & Maximum speed (standardized) & 520 & 0.104 & 0.556 \\
Segments & Signalized segment & 673 & 0.135 & 0.746 \\
Segments & Overdispersion $\kappa$ & 5000 & 1.000 & 0.017 \\
Intersections & Intercept & 334 & 0.067 & 0.842 \\
Intersections & Four or more incident segments & 243 & 0.049 & 0.897 \\
Intersections & Highest incident speed & 259 & 0.052 & 0.896 \\
Intersections & Second-highest incident speed & 240 & 0.048 & 0.898 \\
Intersections & Land-use treatment: Renewal & 308 & 0.062 & 0.871 \\
Intersections & Land-use treatment: Conservation & 237 & 0.047 & 0.910 \\
Intersections & Land-use treatment: Development & 46 & 0.009 & 0.983 \\
Intersections & Land-use treatment: Integral improvement & 51 & 0.010 & 0.980 \\
Intersections & Signalized-access intensity & 502 & 0.100 & 0.757 \\
Intersections & Overdispersion $\kappa$ & 5000 & 1.000 & 0.005 \\
\bottomrule
\end{tabular}%
}
\TableCaptionBottom{tab:appendix_mcmc_diagnostics}{Effective sample size and lag-one autocorrelation of the retained posterior draws.}
\end{table}

The overdispersion parameters mixed well, but several regression coefficients had low
effective sample sizes and high lag-one autocorrelation. These results justify the
cautious treatment of pavement-surface and less frequent land-use contrasts in the main
text. The deviance-based summaries were $p_D=10.560$ and DIC $=48{,}355.930$ for
segments, and $p_D=9.393$ and DIC $=33{,}989.190$ for intersections. Because these
values correspond to different observational supports, they are reported as numerical
diagnostics rather than used to rank the two components.

\section{Additional spatial diagnostics}
\label{app:spatial_diagnostics}

The following figures are retained as diagnostic complements to the main spatial results. They are useful for reviewing raw-scale discrepancies, but the main article emphasizes the log-scale maps and scatter plot because those displays reduce the influence of extreme counts.

\begin{figure}[H]
\centering
\SafeIncludeGraphics[width=0.92\textwidth]{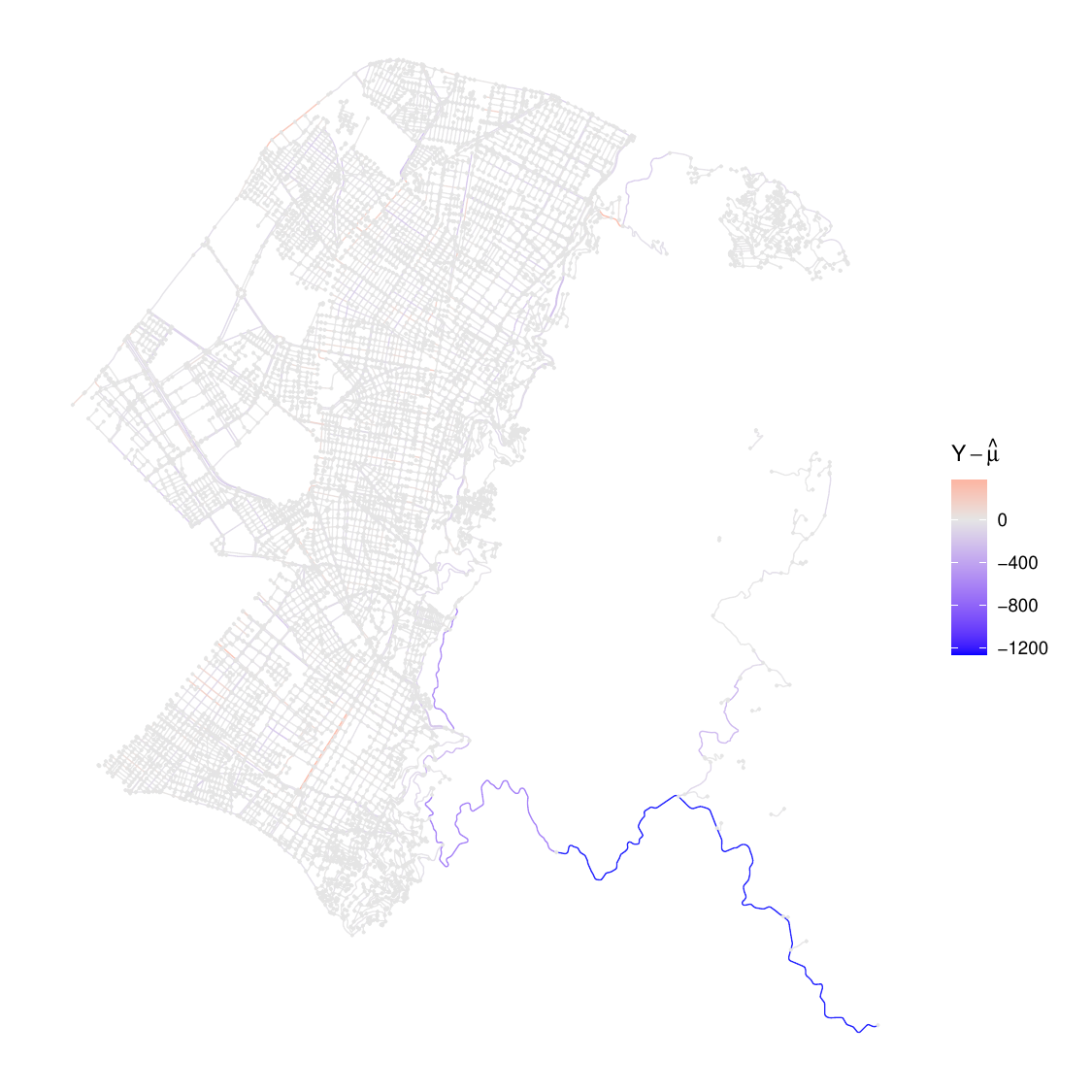}
\caption{Raw residual map, defined as observed counts minus posterior mean counts.}
\label{fig:raw_residual_map}
\end{figure}

\begin{figure}[H]
\centering
\SafeIncludeGraphics[width=0.82\textwidth]{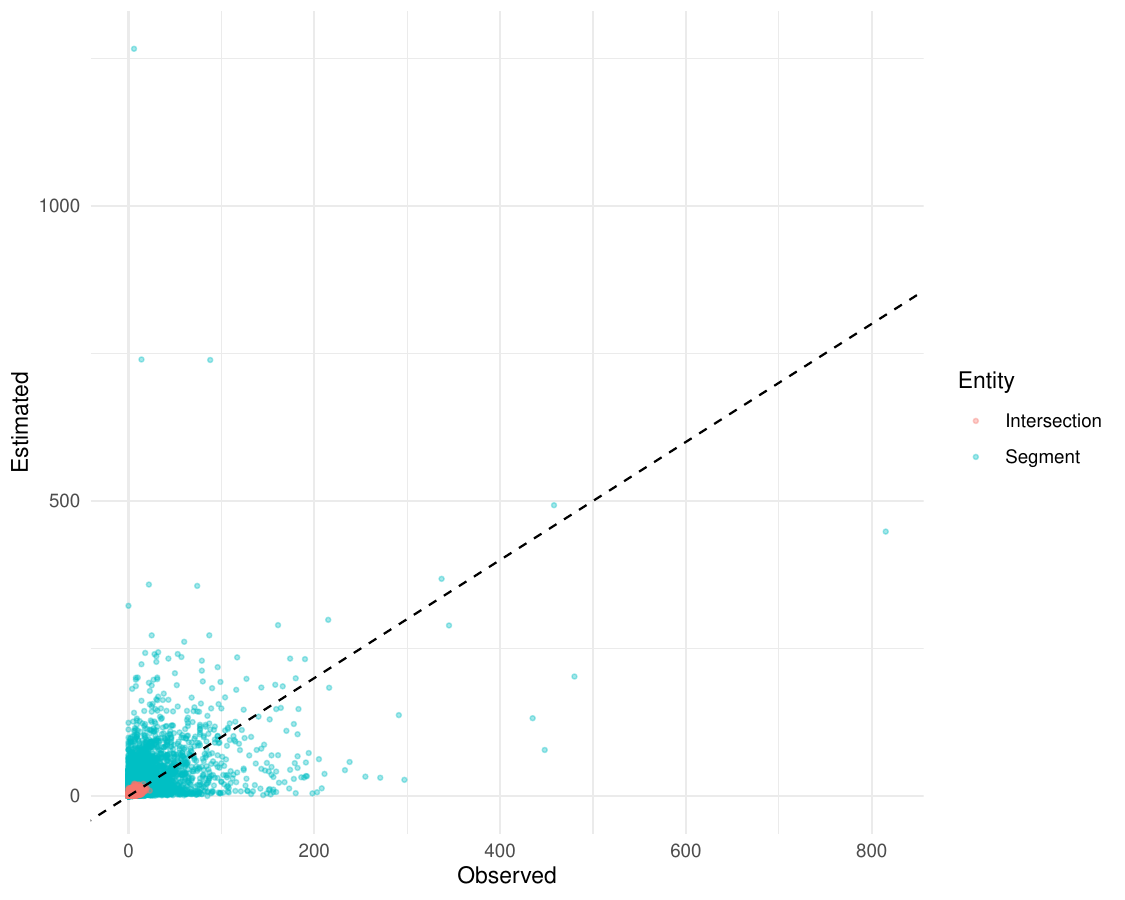}
\caption{Observed versus estimated crash counts in the original count scale.}
\label{fig:raw_scatter}
\end{figure}

\end{document}